\documentclass[twocolumn,showpacs,preprintnumbers,amsmath,amssymb]{revtex4}
\input{epsf.tex}

\def\be{\begin{equation}}
\def\ee{\end{equation}}
\def\ba{\begin{array}}
\def\ea{\end{array}}
\def\bea{\begin{eqnarray}}
\def\eea{\end{eqnarray}}
\begin{document}
\draft
\title {Mass Dependence of Disappearance of Transverse Flow}
\author {Aman D. Sood and Rajeev K. Puri\\
\it Physics Department, Panjab University, Chandigarh -160 014, India\\}
\begin{abstract}
A complete theoretical study is presented for the disappearance
of flow, for the first time, by analyzing 15 reactions
with masses between 47 and 476 units. We demonstrate that the effect of
nucleon-nucleon cross-section reduces to insignificant level for heavier
colliding nuclei in agreement with previous studies. A
stiff equation of state with nucleon-nucleon cross-sections $\sigma$=35-40 mb is able to explain
all the measured balance energies within few percent. A power law
($\propto$$A^{\tau}$) is also given for the mass dependence of the disappearance
of flow which is in excellent agreement with experimental data.
\end{abstract}
\maketitle
\section{Introduction}
The heavy-ion collisions at intermediate energies provide
a rich physical insight into the reaction dynamics. One has measured (and/or
predicted) several new phenomena that may shed light on the nature of hot and
dense nuclear matter formed during a collision. In addition, one also hopes to
understand the nature of nuclear interactions in medium. The prediction of
collective transverse flow by the hydrodynamical model was a very important step
towards the understanding of excited nuclear matter \cite {scheid74}. The collective
transverse flow was found to be very sensitive towards different signals of
excited nuclear matter.
Apart from the transverse flow, one has also proposed
e.g. differential flow \cite {li99}, and elliptic flow \cite {zheng99}. All these
quantities are assumed to be sensitive towards the (nuclear matter) equation of
state and/or nucleon-nucleon (nn) cross-section$\Rightarrow$
{\it the ultimate goal of the intermediate energy heavy-ion collisions}.
One should however, keep in the mind that the reaction
dynamics depends also on the incident energy as well as on the impact parameter of
the reaction. At low incident energies, the dynamics is governed by the attractive
mean field whereas the repulsive interactions decide the fate of a reaction
at higher incident energies. Naturally, the effect of
nn collisions decreases with decrease in the incident energy.
The dominance of the attractive mean field (at low incident energies)
may prompt the emission of
particles into backward hemisphere whereas if nn scatterings
dominate, the particle emission is likely to be in the forward hemisphere.
Therefore, while going from the low incident energy to higher energy,
the attractive
interactions may be balanced by the repulsive interactions, resulting in the
net zero flow (i.e. the disappearance of flow). The energy at which the flow disappears is termed as the
balance energy \cite {mol85}.

During the last few years, extensive efforts have been
made to measure and understand the disappearance of flow
\cite{west93,pak96,ang97,sull90,krof91,he96,cus02,mag0062,westnpa,krof92,zhang90,mag0061,buta95}.
One has measured
the balance energy ($E_{bal}$) in 
$^{12}$C+$^{12}$C
\cite {west93}, $^{20}$Ne+$^{27}$Al \cite {west93}, $^{36}$Ar+$^{27}$Al
\cite {ang97,buta95}, $^{40}$Ar+$^{27}$Al \cite {sull90}, $^{40}$Ar+$^{45}$Sc
\cite {west93,pak96,mag0062},
$^{40}$Ar+$^{51}$V \cite {krof91}, $^{64}$Zn+$^{27}$Al \cite {he96},
$^{40}$Ar+$^{58}$Ni \cite {cus02}, $^{64}$Zn+$^{48}$Ti \cite {buta95},
$^{58}$Ni+$^{58}$Ni \cite {cus02,mag0062,westnpa}, $^{58}$Fe+$^{58}$Fe \cite{westnpa},
$^{64}$Zn+$^{58}$Ni \cite {buta95},
$^{86}$Kr+$^{93}$Nb \cite {west93,mag0062}, $^{93}$Nb+$^{93}$Nb \cite {krof92},
$^{129}$Xe+$^{118}$Sn \cite {cus02}, $^{139}$La+$^{139}$La \cite {krof92},
and $^{197}$Au+$^{197}$Au \cite {mag0062,zhang90,mag0061} systems.
The very recent and accurate measurement of the balance energy $E_{bal}$ in 
$^{197}$Au+$^{197}$Au \cite {mag0062,mag0061} has generated a renewed interest
in the field \cite{cus02}.
Interestingly, most of the reported reactions are symmetric in nature. It should be
kept in the mind that the reaction dynamics depends also upon the asymmetry of the
reaction \cite {singh00}. All the above mentioned measurements were for the central
collisions only. Some measurements \cite{pak96,sull90,he96,cus02,mag0062,westnpa}
however, also
took the impact parameter dependence into account. As noted in ref.
\cite {mag0062}, the $E_{bal}$ for heavier nuclei shows a little dependence on
the impact parameter whereas a large variation in the $E_{bal}$
can be seen for lighter colliding nuclei \cite {pak96,mag0062}.
The possible cause could be the fact that 
the disappearance of flow for heavier nuclei occurs at a much lower incident
energy compared to lighter nuclei
(e.g. the measured $E_{bal}$ for $^{197}$Au+$^{197}$Au is 42$\pm$3$\pm$1
MeV/nucleon \cite{mag0061} whereas it is 111$\pm$10 MeV/nucleon
for $^{20}$Ne+$^{27}$Al \cite{west93}). In the (nearly) absence of nn collisions at low
incident energies,
a little dependence should occur on impact parameter \cite {puri98}.
Some attempts also exist in the literature where enhancement in the
$E_{bal}$ with neutron content was found experimentally/theoretically
\cite {westnpa,lie98,li96}.

The above findings reveal the measurements of balance
energy in more than 15 systems ranging from very light to heavy nuclei. As a result,
a power law behaviour ($\propto$$A^{\tau}$) has also been reported for 
$E_{bal}$ \cite {west93,mag0062,mag0061,buta95}.
Earlier the power law parameter ``$\tau$'' was supposed
to be close to --1/3 (resulting from the interplay between the
attractive mean field
and repulsive nn collisions) \cite {west93} whereas the recent
measurements
suggest a deviation from the above mentioned power law \cite {mag0062,mag0061}.

Various theoretical attempts have been made to understand the vanishing of
nuclear flow. Most of these are, however, using the Boltzmann-Uehling-Uhlenbeck
model \cite{li99,zheng99,mol85,west93,sull90,he96,mag0062,westnpa,krof92,mag0061,li96,zhounpa,kla93,xu92,li93,zhou94}.
Some
attempts are also reported in the literature where Quantum Molecular Dynamics
(QMD) model was used \cite {pak96,lie98,soff95,leh96,suneel98,soodsym,sood03}.
Different theoretical
attempts considered either a stiff or soft equation of state along with variety of
nn cross-sections. Interestingly, out of all these studies,
only a couple of attempts exist where mass dependence
of the disappearance of flow was discussed
\cite {west93,mag0062,mag0061,buta95,zhou94,soodsym,sood03}.
A careful analysis shows that in refs. \cite {west93,zhou94}
a total mass $\leq$200 was considered for the power law studies
whereas in refs. \cite {soodsym,sood03} only
heavier masses $\geq$175 were analyzed.
There the $E_{bal}$ for heavier nuclei was found to scale as approximately
$1/\sqrt {A}$ \cite {soodsym,sood03} whereas lighter and medium mass nuclei
follow A$^{-{1/3}}$ dependence \cite{west93,zhou94}.
Refs. \cite {mag0062,mag0061}
included for the first time, a larger mass range 63/47$\leq$A$\leq$394. However, even
in these studies, only six systems were taken into account (note that the
power law fit was made for those systems which were recorded in the 4$\pi$ NSCL
experiment. The data from the GANIL experiments, however,
was not taken into account for fitting).
The earlier mass dependence by the same group \cite {west93} also considered five systems only. We
here present, for the first time, a complete study of the mass dependence in the disappearance of flow
by analyzing as many as fifteen central collisions ranging from $^{20}$Ne+$^{27}$Al
to $^{197}$Au+$^{197}$Au/$^{238}$U+$^{238}$U. The corresponding experimental measurements were
reported by several different groups
\cite{west93,pak96,ang97,sull90,krof91,he96,cus02,mag0062,westnpa,krof92,zhang90,mag0061,buta95}.
This study, therefore, will present a
unified analysis of the balance energy irrespective of the experimental
setup/group. The present study is carried out within the framework of the
Quantum Molecular Dynamics (QMD) model
\cite{singh00,puri98,lie98,soff95,leh96,suneel98,soodsym,sood03,suneel981,aich,hart}.
The details of the model are given
in section 2. The results and discussion is presented in section 3 and we
summarize our results in section 4.\

\section{The Model}
We describe the time evolution of a heavy-ion reaction within the framework of Quantum
Molecular Dynamics (QMD) model
\cite{singh00,puri98,lie98,soff95,leh96,suneel98,soodsym,sood03,suneel981,aich,hart}
which is based on a molecular dynamics picture. 
Here each nucleon is represented by a coherent state of the form
\begin{equation}
\phi_{\alpha}(x_1,t)=\left({\frac {2}{L \pi}}\right)^{\frac {3}{4}}
e^{-(x_1-x_{\alpha }(t))^2} e^{ip_{\alpha}(x_1-x_{\alpha})}
e^{-\frac {i p_{\alpha}^2 t}{2m}}.  \label {e1}
\end{equation}
Thus, the wave function has two time dependent parameters
$x_{\alpha}$ and $p_{\alpha}$.  The total n-body wave function
is assumed to be a direct product of coherent states:
\begin{equation}
\phi=\phi_{\alpha} (x_1,x_{\alpha},p_{\alpha},t)\phi_{\beta}(x_2,x_{\beta},
p_{\beta},t)....,         \label {e2}
\end{equation}
where, antisymmetrization is neglected. The initial values of the parameters
are chosen in a way that the ensemble ($A_T$+$A_P$)  nucleons give a proper density
distribution as well as a proper momentum distribution of the projectile and
target nuclei. The time evolution of the system is calculated using the
generalized variational principle. We start out from the action
\begin{equation}
S=\int_{t_1}^{t_2} {\cal {L}} [\phi,\phi^{*}] d\tau, \label {e3}
\end{equation}
with the Lagrange functional 
\begin{equation}
{\cal {L}} =\left(\phi\left|i\hbar \frac {d}{dt}-H\right|\phi\right), \label {e4}
\end{equation}
where the total time derivative includes the derivatives with respect to the
parameters. The time evolution is obtained by the requirement that the action is
stationary under the allowed variation of the wave function
\begin{equation}
\delta S=\delta \int_{t_1}^{t_2} {\cal {L}} [\phi ,\phi^{*}] dt=0.
\label{e5}
\end{equation}
If the true solution of the Schr\"odinger equation is contained in the
restricted set of wave function
$\phi_{\alpha}\left({x_{1},x_{\alpha},p_{\alpha}}\right),$
this variation of the action gives the exact solution
of the Schr\"odinger equation. If the parameter space is too restricted, we obtain
that wave function in the restricted parameter space which comes close to the
solution of the Schr\"odinger equation.
Performing the variation with the test wave function (2), we obtain for each
parameter $\lambda$ an Euler-Lagrange equation;
\begin{equation}
\frac{d}{dt} \frac{\partial {\cal {L}}}{\partial {\dot {\lambda}}}-\frac{\partial \cal {L}}
{\partial \lambda}=0.  \label{e6}
\end{equation}
For each coherent state and a Hamiltonian of the form,
$H=\sum_{\alpha} \left[T_{\alpha}+{\frac{1}{2}}\sum_{\alpha\beta}V_{\alpha\beta}\right]$,
the Lagrangian and the Euler-Lagrange function can be easily calculated \cite{aich}
\begin{equation}
{\cal {L}} = \sum_{\alpha}{\dot {\bf x}_{\alpha}} {\bf p}_{\alpha}-\sum_{\beta}
\langle{V_{\alpha \beta}}\rangle-\frac{3}{2Lm}, \label{e7}
\end{equation}
\begin{equation}
{\dot {\bf x}_{\alpha}}=\frac{{\bf p}_\alpha}{m}+\nabla_{p_{\alpha}}\sum_{\beta}
\langle{V_{\alpha \beta}}\rangle, \label {e8}
\end{equation}
\begin{equation}
{\dot {\bf p}_{\alpha}}=-\nabla_{{\bf x}_{\alpha}}\sum_{\beta}
\langle{V_{\alpha \beta}}\rangle. \label {e9}
\end{equation}
Thus, the variational approach has reduced the n-body Schr\"ondinger
equation to a set of 6n-different equations for the parameters which can be
solved numerically. If one inspects  the formalism carefully, one finds that the
interaction potential which is actually the Bruckner G-matrix
can be divided into two parts: (i) a real part and (ii) an imaginary part.
The real part
of the potential acts like a potential whereas the imaginary
part is proportional to the cross-section.

In the present model, interaction potential comprises of the following terms:
\begin{equation}
V_{\alpha\beta} = V_{loc}^{2} + V_{loc}^{3} + V_{Coul} + V_{Yuk},
\label {e10}
\end {equation}
$V_{loc}$ is the Skyrme force whereas $V_{Coul}$ and $V_{Yuk}$ define,
respectively, the Coulomb and Yukawa terms.
The expectation value of  these potentials is  calculated as
\begin{eqnarray}
V^3_{loc}& =& \int f_{\alpha} ({\bf p}_{\alpha}, {\bf r}_{\alpha}, t)
f_{\beta}({\bf p}_{\beta}, {\bf r}_{\beta}, t)V_I ^{(2)}({\bf r}_{\alpha},
{\bf r}_{\beta})
\nonumber\\
&  & \times
{d^{3} {\bf r}_{\alpha} d^{3}
{\bf r}_{\beta}  d^{3}{\bf p}_{\alpha}  d^{3}{\bf p}_{\beta},}
\end{eqnarray}
\begin{eqnarray}
V^3_{loc}& =& \int  f_{\alpha} ({\bf p}_{\alpha}, {\bf r}_{\alpha}, t)
f_{\beta}({\bf p}_{\beta},
{\bf r}_{\beta},t) f_{\gamma} ({\bf p}_{\gamma}, {\bf r}_{\gamma}, t)
\nonumber\\
&  & \times  V_I^{(3)} ({\bf r}_{\alpha},{\bf r}_{\beta},{\bf r}_{\gamma})
d^{3} {\bf r}_{\alpha} d^{3} {\bf r}_{\beta} d^{3} {\bf r}_{\gamma}
\nonumber\\
&  & \times d^{3} {\bf p}_{\alpha}d^{3} {\bf p}_{\beta}
d^{3} {\bf p}_{\gamma}.
\end{eqnarray}
Where $f_{\alpha}({\bf p}_{\alpha}, {\bf r}_{\alpha}, t)$ is the Wigner density which
corresponds to the wave functions (eq. 2).
If we deal with the local Skyrme force only, we get
\small{\begin{equation}
V^{Skyrme} = \sum_{{\alpha}=1}^{A_T+A_P} \left[\frac {A}{2} \sum_{{\beta}=1}
\left(\frac {\tilde{\rho}_{\alpha \beta}}{\rho_0}\right) +
\frac {B}{C+1}\sum_{{\beta}\ne {\alpha}}
\left(\frac {\tilde {\rho}_{\alpha \beta}} {\rho_0}\right)^C\right].
\end{equation}}

\normalsize
Here A, B and C are the Skyrme parameters which are
defined according to the ground state properties of a nucleus.
Different values of C lead to different equations
of state. A larger value of C (=380 MeV) is often dubbed as stiff equation of state.

A number of attempts exist in the literature
which study the nature of equation of state. Following refs.
\cite {li99,zheng99,mol85,pak96,sull90,he96,zhounpa,xu92,zhou94,soff95,leh96,suneel98,soodsym,sood03},
we shall also employ a stiff equation of state through
out the present analysis.
It should also be noted that the success rate is nearly the same for stiff
and soft equations of state.
Further, it has been
shown in refs. \cite {sull90,he96,zhounpa,xu92,zhou94} that the difference
between $E_{bal}$ using a stiff and soft equation of state is
insignificant for central heavy-ion collisions.

The imaginary part of the
potential i.e. the nn cross-section has been a point of
controversy. A large number of calculations exist where a constant
and isotropic cross-section is used. Following refs.
\cite {sull90,he96,zhounpa,xu92,suneel98,soodsym,sood03,barz96,roy97},
we also use constant energy independent cross-section. As 
shown by Li \cite{li93}, most of the collisions below 100 MeV/nucleon
happen with nn cross-section of 55 mb strength. Keeping the present energy
domain into mind, the choice of a constant cross-section is justified.
It has also been also shown by Zheng et al. \cite {zheng99} that
a stiff equation of state with free nn cross-section and a soft
equation of state with reduced cross-section yield nearly the same results.
For comparison, we shall also use an energy dependent cross-section as fitted
by Cugnon \cite{aich} (labeled as Cug) as well as a medium dependent
cross-section derived from G-matrix \cite{aich} (denoted by GMC).

\section{Results and Discussion} Using a stiff equation of state along with
different nn cross-sections, we simulated the above mentioned reactions for
1000-3000 events in each case. The reactions were followed till transverse
flow saturates. The saturation time varies between 150 fm/c
(for lighter colliding nuclei such as $^{20}$Ne+$^{27}$Al) and 300 fm/c
(for heavier colliding nuclei such as $^{197}$Au+$^{197}$Au). In particular,
we simulated
$^{20}$Ne+$^{27}$Al at b=2.6103 fm,
$^{36}$Ar+$^{27}$Al at b=2 fm, $^{40}$Ar+$^{27}$Al at b=1.6 fm,
$^{40}$Ar+$^{45}$Sc
at b=3.187 fm, $^{40}$Ar+$^{51}$V at b=2.442 fm, $^{64}$Zn+$^{27}$Al
at b=2.5 fm, $^{40}$Ar+$^{58}$Ni at b=0-3 fm, $^{64}$Zn+$^{48}$Ti at b=2 fm,
$^{58}$Ni+$^{58}$Ni at b=2.48 fm, $^{64}$Zn+$^{58}$Ni at b=2 fm,
$^{86}$Kr+$^{93}$Nb at b=4.07 fm, $^{93}$Nb+$^{93}$Nb at b=3.104 fm,
$^{129}$Xe+$^{Nat}$Sn at b=0-3 fm, $^{139}$La+$^{139}$La at b=3.549 fm and
$^{197}$Au+$^{197}$Au at b=2.5 fm. The
choice of impact parameter is based on the experimentally extracted information
[5-17]. The above reactions were simulated at incident energies
between 30 MeV/nucleon and 150 MeV/nucleon depending upon the mass of the
system. Naturally, a lower energy range was used for heavy nuclei
whereas a higher beam energy was needed for lighter cases. The reactions
were simulated at different fixed incident energies and a straight line 
interpolation was used to find the balance energy $E_{bal}$.

There are several methods used in the literature
to define the nuclear transverse flow. In most of
the studies, balance energy is extracted from ($p_{x}/A$) plots where
one plots ($p_{x}/A$) as a function of $Y_{cm}/Y_{beam}$. Using a linear fit to
the slope, one can define the so-called reduced flow (F).
Naturally, the energy at which the reduced flow passes through zero, is called 
the balance energy. Alternatively, one can also use a more integrated quantity ``directed
transverse momentum $\langle{p_{x}^{dir}}\rangle$'' which is defined as
\cite {leh96,suneel98,soodsym,sood03,suneel981,aich,hart}:
\begin{equation}
\langle p_{x}^{dir}\rangle~=~\frac{1}{A}\sum_i sign\{Y(i)\}~p_{x}(i),
\end{equation}
where $Y(i)$ and $p_{x}(i)$ are the rapidity distribution and transverse
momentum of the $ith$ particle. In this definition, all rapidity bins
are taken into account. It, therefore, presents an easier way of measuring the
in-plane flow rather than complicated functions such as the ($p_{x}/A$) plots.
It is worth mentioning that the balance energy is independent of the
nature of emitted particle \cite {west93}.
Further, the apparatus correction and acceptance
does not play any role in calculating the energy of vanishing
flow \cite{west93,krof91}.

In fig. 1, we display the final state transverse momentum
$\langle{p_{x}/A}\rangle$ as a function of the rapidity which is defined
as:
\begin{equation}
Y(i) = \frac{1}{2}\ln\frac {{\bf{E}}(i)+{\bf{p}}_{z}(i)}
{{\bf{E}}(i)-{\bf{p}}_{z}(i)},
\end{equation}
where ${\bf E}(i)$ and ${\bf p}_{z}(i)$ are, respectively, the total
energy and longitudinal momentum of $ith$ particle. The upper parts are
for $^{20}$Ne+$^{27}$Al (at 150 fm/c) and $^{40}$Ar+$^{45}$Sc (at 150 fm/c),
whereas the bottom parts are for $^{139}$La+$^{139}$La (at 300 fm/c) and
$^{197}$Au+$^{197}$Au (at 300 fm/c). The middle part is for $^{64}$Zn+$^{58}$
Ni (at 300 fm/c) and $^{93}$Nb+$^{93}$Nb (at 300 fm/c).
In all cases, 
the slope is negative at lower
incident energies
which changes to positive value at higher incident energies. Between these
limits, the slope becomes almost zero at a particular energy.
This zero slope energy is termed as balance energy. One also notices that a
higher value of incident energy is needed in lighter cases to balance the
attractive and repulsive forces. This energy decreases with increase in the
mass of the system.

A look at fig. 2, where
$\langle {p_{x}^{dir}} \rangle$ (instead
of $\langle {p_{x}/A} \rangle$) is plotted, depicts quite similar trends.
Here $\langle {p_{x}^{dir}} \rangle$ is displayed as a function of the
reaction time. The
$\langle {p_{x}^{dir}} \rangle$ during initial stage is always negative
irrespective of the incident energy. This
shows that the interactions among nuclei are attractive during
the initial phase of the 
reaction. These interactions remain either attractive throughout the time
evolution, or may turn repulsive depending on the incident energy. The transverse
flow in lighter colliding nuclei saturates earlier compared to heavy colliding
nuclei. One also sees a sharp transition from negative to positive flow
in lighter nuclei. This transition is gradual when one analyzes the
heavier nuclei. If one compares figs. 1 and 2, one finds that the disappearance of
flow (where flow is zero) occurs at same incident energies in both cases
showing the
equivalence between ($p_{x}/A$) and $\langle {p_{x}^{dir}} \rangle$ as far as balance
energy is concerned. The latter quantity is more useful since it is 
stable than former one. These findings are in agreement with
refs. \cite{leh96,suneel98}.

It has been advocated by several authors that the study of disappearance
of flow can shed light on the magnitude of nn
cross-section \cite{zheng99,west93,sull90,he96,mag0062,westnpa,mag0061,zhounpa,xu92,li93,zhou94,leh96,suneel98,soodsym,sood03}.
To check this point with reference to mass dependence,
we show in fig. 3, the final stage $\langle {p_{x}^{dir}} \rangle$
as a function of the incident energy. The left panel is for $^{20}$Ne+$^{27}$Al,
$^{64}$Zn+$^{58}$Ni and $^{139}$La+$^{139}$La, respectively, for top, middle
and bottom whereas top, middle and bottom of right hand side displays the
results, respectively, of $^{40}$Ar+$^{45}$Sc, $^{93}$Nb+$^{93}$Nb and 
$^{197}$Au+$^{197}$Au. The experimental data are displayed by stars whereas
the present results with $\sigma$=55, 40 and 20 mb are shown, respectively,
by solid circles, open squares and solid triangles.
The $\langle {p_{x}^{dir}} \rangle$ obtained with energy dependent
cross-section due to Cugnon (labeled as Cug) and the one that takes the medium into account (i.e. the G-matrix)
are marked by solid diamonds and inverted triangles, respectively. First of
all, we see that the medium effects in nn cross-section do not play any role at
these low incident energies. The results obtained with Cugnon energy dependent
cross-section and G-matrix medium dependent cross-section are roughly the
same for heavier colliding nuclei where reactions are simulated at low incident
energies. Some visible differences, however, can be seen in the case of light
colliding nuclei where incident energy is relatively high.
One also sees a linear enhancement in the nuclear flow with increase in
the incident energy. Further, the role of different cross-sections is consistent
through out the present mass range. The largest
cross-section gives more positive flow which is
followed by the second larger cross-section. Interestingly,
these effects depend on the mass of the system.
If one looks at the
reaction of $^{20}$Ne+$^{27}$Al, one sees that the $E_{bal}$ increases from
89 MeV/nucleon to 
244 MeV/nucleon when nn cross-sections is reduced from 55 mb to 20 mb. Whereas the range of
$E_{bal}$ for the same cross-sections was 47-83 MeV/nucleon for $^{93}$Nb+$^{93}$Nb
reaction. If one goes to still heavier nuclei, $^{197}$Au+$^{197}$Au, the 
range of $E_{bal}$ narrows down to 38-59 MeV/nucleon. In other
words, a reduction in the cross-section by 64\% percent yields a change of 155
MeV/nucleon in the case of $^{20}$Ne+$^{27}$Al reaction, whereas it is only 21 MeV/nucleon
for the case of $^{197}$Au+$^{197}$Au. Similarly looking at the curves of
$\sigma$=55 and 40 mb,
a reduction in the cross-section by 27\% yields a difference of 30
MeV/nucleon in the $E_{bal}$ for $^{20}$Ne+$^{27}$Al reaction whereas nearly 4 MeV
difference exists
for the case of $^{197}$Au+$^{197}$Au reaction. This result, which is in agreement with the
findings of refs. \cite {mag0062,westnpa}, depicts that for heavier colliding nuclei,
the $E_{bal}$ is independent of the cross-section one is choosing.
Further, the standard energy dependent nn
cross-section (Cug) fails to reproduce the observed balance energy in almost all the
cases. However, a constant cross-section of 40 mb strength seems to be closer
to the experimental observed balance energy. This conclusion is supported by
several other groups where a cross-section of 30-40 mb was used to reproduce the
experimental data
\cite {sull90,he96,zhounpa,xu92,li93,zhou94,suneel98,soodsym,sood03,barz96,roy97}.
This will be discussed in detail in the following
paragraphs.

It has been argued in refs. \cite {leh96,blat91} that the flow at any
point during the reaction can be divided into the parts emerging
from the (attractive) mean field potential and (repulsive) nn collisions.
Following \cite {leh96}, we decomposed the
transverse momentum into the contributions created by the mean field and two-
body nn collisions. This extraction, which is made from the simulations of the
QMD model, is done as following \cite {leh96}: Here at each time step during the
collision, momentum transformed by the mean field and two- body collision
is analyzed separately. Naturally, we get two values at each step which
can be followed throughout the reaction. The total transverse momentum can be
obtained by adding both these contributions.

In fig. 4, we display the final state
$\langle {p_{x}^{dir}} \rangle$ decomposed into two parts i.e. into the mean field
and two- body collision parts as a function of the incident energies for different
colliding systems as reported in fig. 3. Again a linear enhancement in the
flow with energy can be seen. Further, the contribution of the mean field
remains attractive throughout the energy range whereas collision contribution
is always repulsive. The balancing of both these contributions results in net
zero flow.
One should, however, keep in the mind that the
contribution of the mean field potential may even
turn repulsive at higher incident energies \cite {blat91}.

In fig. 5, we display the energy of vanishing flow
($E_{bal}$) as a function
of the combined mass of the system for fifteen different
colliding pairs that range from
$^{20}$Ne+$^{27}$Al to $^{197}$Au+$^{197}$Au. In our earlier
communication \cite {sood03}, a prediction of $E_{bal}$
for $^{238}$U+$^{238}$U was also reported. It is worth mentioning that this
is the first ever attempt that deals with as many as fifteen systems. Earlier
attempts of mass dependence \cite {west93,mag0062,mag0061,buta95,zhou94} had taken
about six cases into account. Apart from the experimental data, we also show
our results for $\sigma$=35 and 40 mb. All curves are a fit of the form
$c.A^{\tau}$ using $\chi^{2}$ minimization procedure. The
experimental data can be fitted by $\tau_{expt}=-0.42079\pm0.04594$
whereas our present results
with $\sigma$=40 mb has $\tau_{40}=-0.41540\pm0.08166$.
The results with $\sigma$=35 mb
yields $\tau_{35}=-0.43037\pm0.08558$. Exclusively, one can extract the
following:

\begin{itemize}
\item The present value of $\tau_{expt}$ differs from the
earlier reported results ($\tau_{expt}=-1/3$ for A$\leq$200
\cite {west93} and 
$\tau_{expt}=-0.46\pm0.06$ \cite {mag0061} for A$\leq$400 mass).
Note that in earlier attempts \cite {mag0062,mag0061}, the
data of NSCL alone was used to fit the power law. In the present case,
we have taken data from different sources amounting to 15 reactions. Therefore,
the present analysis is universal in nature.

\item The present theoretical value
$\tau_{40}=-0.41540\pm0.08166$
is the closest
one obtained so far. In earlier reports \cite {west93}, the 
$\tau_{expt}$ was $-1/3$ whereas BUU model
yielded $-0.28$$\leq$$\tau_{th}$$\leq$$-0.32$.
In another study \cite {mag0061}, the $\tau_{expt}$ was
$-0.46\pm0.06$ whereas BUU model has
$\tau_{th}=-0.41\pm0.03$. In other words, the present QMD model
with a stiff EOS along with
$\sigma$=35-40 mb can explain the data much better than any other theoretical
calculations. Our present analysis has much wider mass spectrum than
any early attempt. Some deviations in the middle order are also reported by
other authors \cite{li93}. The $\sigma$=40 explains $E_{bal}$ in heavier
nuclei whereas $\sigma$=35 reproduces a middle order nicely.

\item From the figure, it is also evident that a true cross-section for this
energy domain should be between 35 and 40 mb. This conclusion is very
important since a  wide range of masses was used for the present analysis.
Our conclusion about the strength of the nn cross-section
is in agreement with large number of earlier calculations on disappearance of
flow and other phenomena in heavy-ion collisions
\cite {sull90,he96,zhounpa,xu92,li93,zhou94,suneel98,soodsym,sood03,barz96,roy97}.
As noted in ref.
\cite {zhou94}, this value of nn cross-section is still much less than the actual
averaged free nn cross-section which is about 60 mb \cite
{chen68}. In the case of fragmentation, a larger cross-section of $\sigma$=55 mb
is also suggested \cite {barz96}.
\end{itemize}

We have also tried
to fit the balance energy in terms of other parameters such as the
charge of the colliding
nuclei. This attempt is shown in fig. 6 where $E_{bal}$ is plotted as a
function of the total atomic number of the system. In the upper part, we display
the full range of the systems whereas in the lower part, only heavier nuclei
are taken into picture. A power law $\propto$$Z^{\tau}$
fits the data nicely. Now $\tau$ is $-0.46703\pm0.04745$ ($-0.45808\pm0.08688$)
for experimental data (theoretical results) which is larger compared to mass
power law ($\tau=-0.42079\pm0.04594 (-0.41540\pm0.08166)$) for experiment data (theoretical results).
This difference in the slopes stems from the different charge/mass ratio
in lighter and heavy nuclei.
Interestingly, the value of $\tau$ for heavier nuclei
(see fig. 6(b)) is $-0.59316\pm0.0622$ ($-0.58483\pm0.11409$).
This result, which is in agreement
with ref. \cite {mag0061}, shows the dominance of the Coulomb
interactions in heavier colliding nuclei.
It would be of further interest to investigate whether the flow due to
the collision and mean field parts (at the balance energy) exhibit any
mass dependence or not. We display,
in fig. 7, the flow $\langle p_{x}^{dir} \rangle$ at balance
energy due to the collision
(upper) and mean field parts (lower). Interestingly, we do not see
any clear mass dependence. Rather very weak dependence
(with $\tau=-0.09427\pm0.08379$) exists on the system size.
This is in agreement with ref.
\cite {blat91}, where a similar conclusion was drawn.

\section{Summary} We have studied the mass dependence of the disappearance of flow
in large number of colliding nuclei using QMD model. As many as fifteen
reactions with masses between 47
and 476 were studied for the first time, where experimental balance energy
was available. Our findings suggest a weak dependence of different
cross-sections for heavier colliding nuclei in agreement with \cite {mag0062}. For the
first time, fifteen different reactions were studied and
our calculations with a stiff equation of state are in excellent agreement with
experimental data. We could reproduce the slope of the power law
($\propto$$A^{\tau}$) very well.
Our calculations suggest a cross-section
of 35-40 mb in this incident energy domain. We also showed that the
collective flow due to mean field is attractive whereas it is repulsive for
collision part. The balancing of these two parts results in
the disappearance of flow.\\

{\it This work is supported by the grant (No. SP/S2/K-21/96)
from Department of Science and Technology, Government of India.}
\newpage

\newpage

{\Large \bf Figure Captions}\\

{\bf FIG. 1.}
The averaged $\langle p_{x}/A \rangle$ as a function of $Y_{c.m.}/Y_{beam}$.
Here we display the results at different incident energies using a stiff
equation of state along with $\sigma$=40 mb. The reactions of
$^{20}$Ne+$^{20}$Al and $^{40}$Ar+$^{45}$Sc are at 150 fm/c whereas those of
$^{64}$Zn+$^{58}$Ni, $^{93}$Nb+$^{93}$Nb, $^{139}$La+$^{139}$La and
$^{197}$Au+$^{197}$Au are at 300 fm/c.\\

{\bf FIG. 2.} The time evolution of $\langle p_{x}^{dir} \rangle$ as a
function of time. Here again results are for stiff equation of state along
with $\sigma$=40 mb.\\

{\bf FIG. 3.} The
$\langle p_{x}^{dir} \rangle$ as a function of the incident
energy. The results for different cross-sections of 55, 40 and 
GMC are represented, respectively, by the solid circles, open squares and
solid inverted triangles whereas for Cug and 20 mb are represented by solid
diamonds and solid triangles. A stiff equation of state has been used. All
lines are to guide the eyes.\\

{\bf FIG. 4.} The decomposition of $\langle p_{x}^{dir} \rangle$ into collision
and mean field parts as a function of incident beam energy. Here results are
displayed for $\sigma$=40 mb. Stars are the experimental balance energy.\\

{\bf FIG. 5.} The balance energy as a function of combined mass of the system.
The experimental points along with error bars are displayed by solid stars whereas
our theoretical calculations for $\sigma$=35 and 40 mb are shown by open
triangles and squares.
The lines are the power law $\propto$$c.A^{\tau}$. The solid, dashed and
dash-dotted lines represent the power law fit for experimental points, with
$\sigma$=40 and 35 mb, respectively.\\

{\bf FIG. 6.} (a) Balance energy as a function of atomic number Z. Here we
display the experimental results along with our calculations for $\sigma$=40 mb.
The fits are obtained with $\chi^{2}$ minimization for power law function 
$c.Z^{\tau}$. (b) Same as (a), but for heavier nuclei.\\

{\bf FIG. 7.} The decomposition of the $\langle p_{x}^{dir} \rangle$ at balance
energy into collision and mean field parts. The results are obtained
using a stiff equation of state along with $\sigma$=40 mb.\\

\end{document}